\documentclass[twocolumn,prl,aps]{revtex4}
\usepackage{graphicx}
\begin{document}
\renewcommand{\thefootnote}{\fnsymbol{footnote}}
\sloppy
\newcommand \be{\begin{equation}}
\newcommand \bea{\begin{eqnarray} \nonumber }
\newcommand \ee{\end{equation}}
\newcommand \eea{\end{eqnarray}}
\newcommand{\rar}{\rightarrow}
\newcommand{\eq}{equation}
\newcommand{\eqs}{earthquakes}
\newcommand{\rp}{\right)}
\newcommand{\lp}{\left(}

\title{A Two-Threshold Model for Scaling Laws of Non-Interacting Snow Avalanches}
\author{Jerome Faillettaz }
\affiliation{ Laboratoire 3S, Universite de Grenoble, France}
 \author{Francois Louchet}
\affiliation{ Laboratoire LTPCM,  INPG Grenoble, France}

\author{Jean-Robert Grasso}
\altaffiliation[Now at ]{earthquake hazard team USGS,\\
Menlo Park, California}
\affiliation{LGIT,
Observatoire de Grenoble, France}

\date{\today}

\begin{abstract}
       The sizes of snow slab failure that trigger snow avalanches
       are power-law distributed. Such a power-law probability distribution function
    has also been proposed to characterize different landslide types. In order to understand
     this scaling for gravity driven systems, we introduce a two-threshold 2-
     d cellular automaton, in which failure occurs irreversibly.
  Taking snow slab avalanches as a model system, we find that the sizes of the largest avalanches
   just preceeding the lattice system breakdown are power law distributed.
    By tuning the maximum value of the ratio of the two failure thresholds
    our model reproduces the range of power law exponents observed for land-,
     rock- or snow avalanches. We suggest this control parameter represents
      the material cohesion anisotropy.
\end{abstract}

\maketitle

\vskip 0.5cm

Most natural avalanches, including landslides, rockfalls, turbidites and snow avalanches
exhibit scale-invariant statistics
 \cite{F69,G70,N93,HSA97,PMBT97,MT99,DGH03,RGF94,L02,BL02},
  i.e. obey power laws, $N(s) \sim s^{-b}$,
 where $N(s)$ is the number of events of size $\ge s$.
  Most often different underlying physical mechanisms
  in different experimental conditions may give rise to similar scaling behaviours.
    For the sake of simplicity, the exponents dealt with here are expressed
    in terms of probability distribution functions of areas in order
    to compare the exponents of different types of avalanches.
     Typical $ b$-values drawn from field data are $1.75 \pm 0.3$ for rockfalls
     \cite{DGH03} and  $2.8 \pm 0.5$ for mixed landslides
    \cite{HSA97,PMBT97,MT99,DGH03}.
     Using numerical simulations
      [e.g. \cite{BTW88,B96,H02,NL97,OFC92,S00,J00,VZ98,HN00,DEA98}],
      numerous studies have been undertaken to try to understand the origin of this scale invariance
      and the values of the scaling exponents. These simulations proceed as follows:
       a "load" variable is assigned to each site i of a grid. These variables grow
       through time until one of them exceeds a threshold value. The corresponding site
        becomes unstable, and the load is redistributed to its neighbours in either a conservative
         or a non conservative way. Various relaxations are observed during a single simulation.
         Those models, based on Bak's sand-pile model \cite{BTW88}, reproduce qualitatively the
         observed scaling behaviour. The exponents do not usually agree with observations except if other
         ingredients (e.g. dissipation or stiffness heterogenieties in
         \cite{NL97})
          or some parameter tuning are introduced
          \cite{NL97,OFC92}.

In the case of snow avalanches,
 slab release results from the expansion
 of a "basal crack" along a weak layer parallel to the slope,
  followed by the opening of a "crown crack" across
  the slab depth as suggested by  Fig. 1, (e.g. for a review  \cite{S03}).
   To  measure the size of the snow slab failure that triggers
    the avalanche flow and to avoid any bias linked to the avalanche flow [e.g. as measured by acoustic emission or
    empirical indexes \cite{L02, BL02}],
     we use the length of the crack starting zone (Fig 2).
     The exponent value of  $2.2 \pm 0.1$  we measured for snow slab avalanches scaling is intermediate between
     rockfall and landslide values.
      Owing to their simple geometry, snow slab avalanches may
      be considered as a model system allowing separate treatments
 and sequential combinations of basal and crown failures.
 We designed a two-dimensional cellular automaton on this basis.

The snow slab is modelled by a 2D network of cells that have 2 failure modes.
 The first one simulates the shear failure between the snow slab and the substrate,
  i.e. the emergence of the basal crack.
  The second is the failure between two adjacent
   cells within the snow slab, i.e. the emergence of the crown crack (Fig. 1).
    The proximity to failure of  a cell is defined by a single variable
    $\zeta_i$ proportional to the applied shear stress $\tau_a$, that is initialised to 0.
     The lattice is initially fully intact with cells having a uniform distribution of strength thresholds.
       Periodic boundary conditions are taken in the horizontal direction. During each run, loading
        increments $\Delta \rho$ are scattered uniformly on each cell.
         A cell fails in shear along the basal plane when its  $\zeta_i$ value exceeds
         a threshold value  $\tau_0$, which brings the $\zeta_i$ value to zero. The excess $\zeta_i$
          value (as well as further loading increments) is then equally redistributed onto
           its non-failed first neighbours (i.e. the simulation is conservative).
            This implies that there is no healing process on a broken cell.
 A failure of the second type occurs between a cell $i$ and one of its neighbours $j$ when the difference
$\parallel \zeta_i -  \zeta_j \|$
  exceeds a slab rupture threshold $\sigma_0$.
As a consequence of load redistribution rules, that aim to simulate the slope effect, our model is polarised,
i.e. the $x$ and $y$  directions have different behaviors (Fig. 3, 4).

A peculiarity of our model relative to other avalanche or sand-pile simulations is that,
in agreement with the mechanics of snow slab failures, we introduced a second failure
mode which is controlled by a finite slab strength threshold.
 Previous attempts in using a "two state" model for sand-pile dynamics either focus on flow dynamics
 (e.g. \cite {B95,HN98}) or are dissipative models (e.g.\cite {HN00}). Another difference with previous studies
 that attempted to simulate natural slides is that there is no healing process for broken cells
  in our stress driven system. The system is ineluctably brought to a final instability, defined
  as the stage at which a macroscopic shear failure (labelled $MS$ event in the following) occurs,
  and reaches the system size (Fig. 4). On our simulations we observe that each time an $MS$ event
  occurs there is a remaining "$C$ cluster" within the MS event cluster made of cells that are
  still unbroken in terms of slab cell interactions, i.e. in terms of the second failure mode.
  Following Zapperi et al. for their random fuse model \cite {ZRSV97} and by analogy to the in-situ
  measured crack area patterns (Fig. 1), we choose the size of the "C cluster" to be the relevant parameter
   to measure the size of the simulated snow slab failure area. This  cluster size is that
    of the slab that remains "cohesive" during the final avalanche cascade.
     We ckecked that this measure of the last avalanche size before the breakdown of the system is
       not affected by the finite size of the system. In the simulation it is a proxy for the size
        of the largest avalanche before the final failure. In-situ, it corresponds to the observed
        initial brittle patch of cohesive slab at the onset of the snow avalanche.
         After the avalanche flow, it is mapped by the crown crack length. The $MS$
          event size that is bounded by the finite size of the grid simulates the cascading
          effect of observed snow avalanche sliding induced by the initial snow slab failure.
          The system is reinitialised before each run using a uniform distribution of strength
           thresholds $\sigma_0$ in an interval between   $\Delta \rho$ and the shear threshold  $\tau_0$.
           Picking up the the largest cluster ($C$ cluster size) within the final avalanche for each run
            from thousands of runs leads to power law distributions of snow slab failure sizes (Fig. 5).\\

The sliding dynamics of two non planar surfaces in contact
 result in strain incompatibilities and de-cohesions.
 They cannot be described by shear strain solely (controlled by a shear threshold $\tau_0$),
 and the introduction of another parameter, as for instance a second failure threshold, is justified.
  The above model, that assigns specific thresholds to basal shear and slab cohesion failures,
  appears to be generic of slope failures (e.g. \cite {DEA98}).
The power law exponents given by our automaton can be varied by tuning a single parameter
 of the failure mechanism geometry, defined by $\alpha = max [\sigma_0 / \tau_0]$, i.e. the maximum value of
  the ratio of slab to shear rupture thresholds. This parameter is a possible measure of the cohesive
   anisotropy of the material. By tuning  $\alpha$, the range of observed values for the
   scaling exponents of the various types of gravity-driven failures can be reproduced (Fig. 6).
    It allows an inverse estimation of their respective cohesive anisotropies. $B$-values of $1.75 \pm 0.3$ for rockfalls
    \cite{DGH03}, $2.2 \pm 0.1$ for snow avalanches (Fig. 2) and $2.8 \pm 0.5$ for landslides  \cite{HSA97,PMBT97, MT99, DGH03}
     are reproduced for $\alpha$ values of 0.6-0.9, 0.45-0.55, 0.2-0.5 respectively.
      $\alpha$  values close to unity correspond
     to isotropic materials, suggesting that the more layered the material structure  is,
     the smaller the $\alpha$  value (i.e. the larger $b$-value).
     This is in agreement with the larger b-value and the more layered structural geology reported for landslide
      than for rockfalls (e.g. \cite {DGH03}). The relatively wider range of $\alpha$  values
       for landslides may reflect a larger anisotropy scatter as compared to snow
        slab layer properties, e.g. \cite {Col91}. For the simple geometry of snow avalanches,
         $\alpha < 1$ suggests a slab strength smaller than the basal shear resistance.
          This finding seems at first glance to contradict the general agreement that crown
          crack opening during snow slab failure is a consequence of a relatively easier basal failure.
           This apparent contradiction is however removed by considering the particular type of loading experienced
           by the snow cover, which is essentially in shear mode parallel to the slope.
           Stresses acting on planes perpendicular to the slope, as for instance tensile stresses
           responsible for crown crack opening at the top of the slab, only arise from shear stress
           or shear resistance gradients. This is the reason why in our automaton crown crack opening
            is controlled by the difference in $\zeta$  values between two neighbouring cells.
             These "slab" stresses are thus usually much smaller than shear stresses acting on "basal" planes,
              and so our finding that the corresponding threshold may be smaller than that for basal failure is not
               surprising.

                Increasing the $\alpha$ value above unity corresponds to increasing the
                 slab failure threshold as compared to the shear failure threshold.
                 It progressively inhibits the slab failure, which will never occur for $\alpha > 2$.
                  Accordingly it increases the probability to observe avalanches that are just driven
                   by the shear failure mode leading to extreme events. They correspond to a departure from
                 the power law behavior as evidenced as a peak in the distribution close to the cutoff value (Fig. 6).
               The analysis of the finite size scaling (Fig. 6)
               suggests that the peak may result from a finite size effect
               that leads the system into a supercritical state, e.g. \cite{NAL96}.

Our observations and model suggest that the observed power law distributions
for independant avalanches in natural gravity-driven (i.e. stress-controlled)
 failures may emerge from avalanches in the vicinity of a breakdown point \cite{ZRSV97}
  rather than from self organization: Instead of the successive relaxations recorded
  during a single run of  stationary systems, we used as an output the size of the largest
  avalanche obtained for each run of a system that macroscopically fails.
   It corresponds to field observations
   in which each avalanche flow destroys the corridor or gully ability
    to endure a new event.
   As suggested by several first order transition models (e.g. \cite{S93,NAL96,ZRSV97}, the scaling of the
   largest clusters (the $C$ clusters) before the discontinuity may arise from the variability of either the
   loading values or the distance of the system from the breakdown for each of the
   largest avalanches. A detailed account of the results of
    this model as well as a complete discussion of these properties, will be reported elsewhere \cite{FLG}.

Introducing a 2-threshold irreversible failure model,
 we can drive the system toward two types of behavior through
  variations of the ratio between these 2 thresholds. For $\alpha < 1$
   the sizes of the last avalanche before the system breakdown are power law distributed
  and the exponent varies as a function of  $\alpha$ value.
  It points out the possibly for our model to mimic a critical point for failure.
   For  $\alpha > 1$,
     the second failure mode (i.e. slab) is progressivily inhibited and
      $\alpha$ may play a role similar to the disorder in the random-field Ising model,
       in which there is a transition from small avalanches to a single spanning avalanche
        (yielding the peak in the distribution), e.g. \cite{PRV2003} and reference therein.
     For all $\alpha <1 $,
    our rupture model suggests that standard statistical physics models
     for rupture (e.g. first order  \cite{ZRSV97}
    or continuous phase transitions \cite{SA98}) may possibly apply
     to natural avalanche phenomena as well as the sand-pile simulations (e.g. \cite{BTW88}).


{}

\newpage
\begin{figure}
\includegraphics[width=7cm]{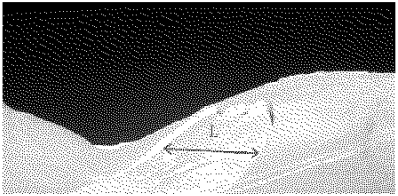}
 \caption{
Snow slab failure which is the avalanche starting zone and the measured $L$ crack length.
 The slab release area is taken as $L^2$. The scale is given by the ski tracks that cross
 the bottom right of the pictures. photograph by A. Caplain.}
\label{fig1}
\end{figure}

    \begin{figure}
\includegraphics[width=7cm]{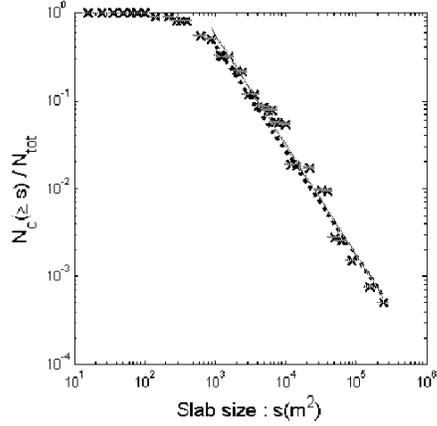}
 \caption{Cumulative distributions of slab released areas obtained from 3935
 blast triggered avalanches (Grande Plagne ski resort, France). The power law is recovered for natural
 avalanche size on the same site with a possible decrease in the exponent value \cite{Fai03}.
The plateau at low sizes results from non-completeness of reports for smaller avalanches than for larger ones.
 Exponent values are $1.2 \pm 0.1$,
  as estimated by the maximum likelihood method \cite{A65} for slab sizes larger than $900$ $m^2$.}
\label{fig2}
\end{figure}

\begin{figure}
\includegraphics[width=4cm]{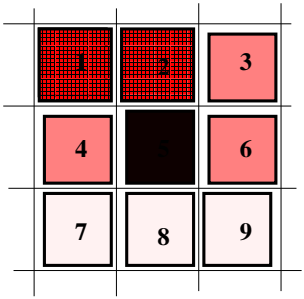}
\caption{Slab failure rules:
($x$  horizontal axis, $y$ vertical axis).
Assume that the cell $\#5$
   is broken in the shear stress mode, and that load redistribution
    on the unbroken "upward" neighbours has occured (grey cells 1, 2, 3, 4, 6).
   Then the slab failure criterion is analysed between cell $\#5$ and
   these 5 upward neighbours which are located in the same $x$ row or in higher $y$ values.
    If the slab failure criterion is fulfilled for a couple of cells (e.g. stripped grey cells),
     the corresponding bonds between these cells break,
     and these cells are no longer considered as neighbours.}
\label{fig3}
\end{figure}

      \begin{figure}
\includegraphics[width=8cm]{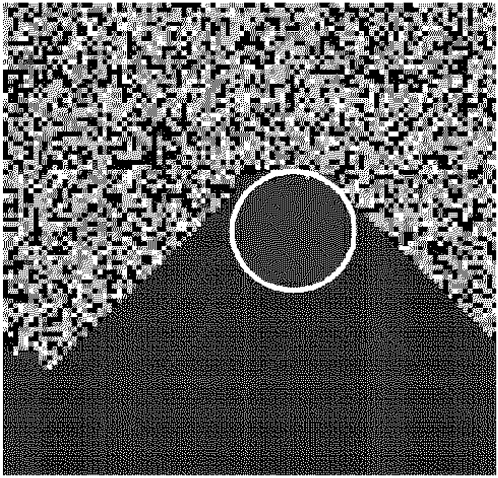}
\caption{   Example of a simulated avalanche for 100 x 100 cells,  ($x$  horizontal axis, $y$ vertical axis).
Red cells (dark grey) represent shear failure between cells and substrate,
and black dots intercell failures within the snow slab.
During loading, some red clusters may appear, until a macroscopic cluster (labelled as $MS$ event)
 suddenly forms,
 extends downslope ($//$ to  $y$ axis)  and reaches the limit of the grid.
  Cells within this red and black $MS$ event have broken in both modes,
   shear failure and lateral cell failure. The $C$ cluster (white circle)
    that remains within the $MS$ event corresponds to a cluster of cells
     where shear failures have occured but not slab cell failures.
      For each simulation the $C$ size is the measured
       output considered in statistics analysis as a proxy to simulate the avalanche starting zone.
      The $45\circ$ pattern of the MS cluster results from the load redistribution rules after
      a shear failures (see Fig. 3). The excess load on a cell at the lower boundary
      of a $C$ cluster cannot be redistributed on its neighbours belonging to the cluster,
       that are already broken; they are therefore mainly redistributed on the 3 neighbour cells
       lying on the row just below.}
\label{fig4}
\end{figure}

                  \begin{figure}
\includegraphics[width=8cm]{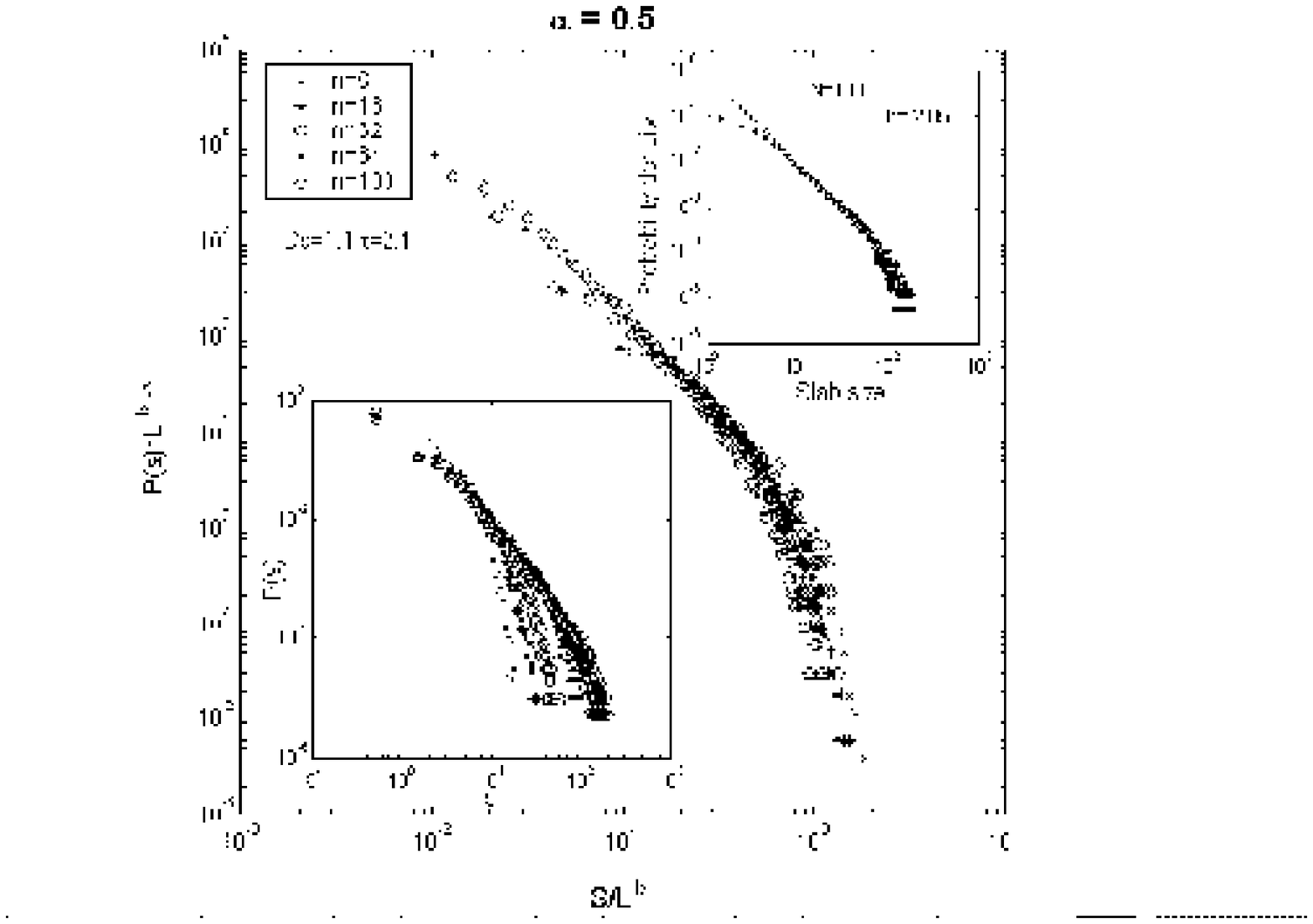}
\caption{  Probality density of simulated avalanche size $S$ for different system sizes, as
measured by the C cluster sizes (starting zone areas), see text for details.
  Load increment values are $\tau_0/4$, where $\tau_0$ is the shear threshold.
For each run  the ratio $\alpha$ of the slab cohesion and shear rupture thresholds
is chosen randomly in a uniform distribution between
a minimum value corresponding to the load increment and
a maximum value $\alpha = max [\sigma_0 / \tau_0]$. Taking $\alpha = 0.5$ gives a $b$-value of 2.05
for the linear part of the plot (solid line, in the 5-100 size range, upper right inset),
in agreement with the experimental results for snow avalanches (Fig. 2).
 The cutoff at large sizes shifts toward larger scales as the grid sizes increases,
 without changing the slope of the linear part of the plot (lower left inset). The finite scaling exponent used for
 central plot are Ds=1.1, $\tau = 2.1$.}
\label{fig5}
\end{figure}

        \begin{figure}
\includegraphics[width=8cm]{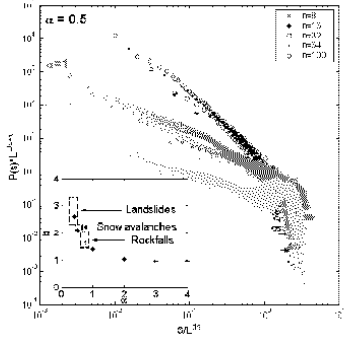}
\caption{Simulated avalanche size distribution as a function of $\alpha$ value.
 Same as on Fig. 5, see text for details. Insert is
the evolution of the exponent of the avalanche size distribution ($b$-value)
as a function of the $\alpha$ values. We isolate two regimes in our model which correspond to the activation
 of either both failure modes (shear and slab failure, $\alpha < 1$) or a single one
 (shear, $\alpha > 2$).
 The exponent value is a function of alpha in the first regime
  and is a constant in the second regime. For $1 < \alpha < 2$
   a crossover between the two regimes leads to a spurious dependence of the b exponent
    on the alpha parameter.
 The finite scaling exponent used for
 central plot are Ds=1.1, $\tau = 2.1$,  Ds=1.3, $\tau = 1.4$,  Ds=1.4, $\tau = 1.$, for $\alpha$ values of 0.5, 1, 2
 respectively.}
\label{fig6}
\end{figure}

\end{document}